\documentclass[prc,twocolumn]{revtex4}

\usepackage{graphicx}
\usepackage{dcolumn}
\usepackage{bm}
\usepackage{amsmath}

\begin{document}

\title{Mean-field calculations of charge radii in ground and isomeric states of 
Cd isotopes}
\author{P.~Sarriguren}
\email{p.sarriguren@csic.es}
\affiliation{
Instituto de Estructura de la Materia, IEM-CSIC, Serrano
123, E-28006 Madrid, Spain}

\date{\today}

\begin{abstract}
Quadrupole moments and charge radii in cadmium isotopes  are studied from a microscopic
perspective. The results obtained from self-consistent deformed Hartree-Fock+BCS 
calculations with Skyrme forces  are compared with isomer and isotope shifts measured 
from high resolution laser spectroscopy experiments. The microscopic calculations 
reproduce fairly well the main features observed in those isotopes that include the 
linear increase with the neutron number of the quadrupole moments of the $11/2^-$ 
isomers, as well as the parabolic behavior of the mean-square charge radii difference 
between isomers and ground states.

\end{abstract}

\maketitle

\section{Introduction}

The theoretical and experimental study of the isotopic evolution of ground and 
low-lying state properties in nuclei has revealed over the years to be a very 
sensitive tool to study highly topical issues in nuclear structure physics, such 
as shell evolution, shape coexistence phenomena or shape transitions
\cite{cheal,campbell,wood-heyde,bender_03}. 
In particular, charge radii and electromagnetic moments 
are very suitable quantities to learn about the nuclear structure properties.
A great experimental effort has been devoted to determine the nuclear charge 
radii. A review of the state of knowledge can be found in Ref. \cite{angeli}, where
different types of experimental data have been compiled. These include absolute
radii measured by muonic-atom spectra and electron scattering experiments, as 
well as relative radii determined from isotope shifts.
The same is true for the nuclear electromagnetic moments that can be found, for 
example, in the tabulation of Ref. \cite{stone} and references therein.

Cadmium isotopes in the neutron mid-shell region have received increasing attention as 
examples where intruder excitations appear.
In fact, the entire  region around Z=50 has been subject 
of discussions \cite{garrett_10,heyde_11,heyde_04} concerning the origin of the 
low-lying excitations.
At first, even-even Cd isotopes were considered clear examples of vibrational 
nuclei, but this picture was later questioned by the presence of intruder 
states opening the possibility of shape coexistence phenomena.
Models that include mixing between 
intruders and phonon states were used to explain the spectroscopic properties 
of these isotopes. Detailed spectroscopic investigations carried out more
recently (see Ref. \cite{garrett_10} for a review) questioned the robustness of 
the spherical vibrator 
picture for these nuclei, suggesting to be deformed $\gamma$-soft rotors.
The issue of the nature of these excitations remains an open problem.

Recently, high precision electromagnetic moments and charge radii have been measured 
in Cd isotopes ($Z=48$) with the collinear laser spectroscopy setup 
at the ISOLDE-CERN radioactive ion beam facility 
\cite{yordanov_13,yordanov_16,yordanov_18,hammen_18}. This is a region of special 
interest because of its proximity to the shell closure at $Z=50$. In 
Ref. \cite{yordanov_13}, quadrupole moments in neutron-rich $A=107-129$ Cd
isotopes were studied. The hyperfine structure was used to identify a spin $1/2^+$
ground-state  in $^{111-119}$Cd isotopes and $3/2^+$ in $^{121-129}$Cd. 
Long-lived isomeric states with spin-parity assignments $11/2^-$ were also firmly 
established in all of these isotopes. The study was extended to lighter odd-$A$ 
Cd isotopes ($A=101-109$ ) in Ref. \cite{yordanov_18}, where ground-state 
spins $5/2^+$ were assigned to them and electromagnetic moments were measured. 
The most remarkable finding in those works was a linear increase of the quadrupole 
moments of the isomeric $11/2^-$ states with the number of neutrons. A plain
explanation of this feature was offered within the extreme shell model, but the
increase was experimentally found well beyond the single $h_{11/2}$ shell. 
The simplicity of the linear increase was investigated microscopically
within the covariant density functional theory  \cite{zhao_14} and it was found 
to be related to the pairing correlations, which smear out the changes induced 
by the single-particle shell structure, leading to a smooth shape evolution. 

In two subsequent papers \cite{yordanov_16,hammen_18} the charge radii of Cd 
isotopes were also studied by high-resolution laser spectroscopy measurements.
In Ref. \cite{hammen_18}, isotope shifts were used to determine the differences 
in mean-square nuclear charge radii of $^{100-130}$Cd isotopes. The extracted 
root-mean-square charge radii show a smooth parabolic behavior on top of a linear 
trend and a regular odd-even staggering. The main features of these measurements 
were not reproduced in detail by standard density functional theory, but were well 
accounted for by a new Fayans functional, which includes a gradient term in the 
pairing functional \cite{reinhard_17} as a distinctive aspect.

In Ref. \cite{yordanov_16} isomer shifts were determined in $^{111-129}$Cd odd-$A$ 
isotopes. Mean-square charge radii differences between the $11/2^-$ isomer states 
and the ground states ($1/2^+$ or $3/2^+$) were measured. They were shown to 
exhibit a parabolic behavior as a function of the neutron number.
This feature was interpreted in terms of a simple model that takes into account 
the existing link between the nuclear radii and the quadrupole deformation of 
the various states. The parabolic behavior arises naturally from the previously 
established \cite{yordanov_13}
linear increase of the deformation parameter in the isomeric $11/2^-$ states
together with the practically constant and small deformation of the ground 
states. The results of this simple model were supported by relativistic mean-field 
calculations.

The purpose of this paper is to study the charge radii, the quadrupole moments, and 
their correlations to see whether those simple properties observed experimentally 
and mentioned above, emerge naturally from microscopic calculations based on
a nonrelativistic self-consistent mean-field formalism with Skyrme effective 
nucleon-nucleon interactions and pairing correlations in the BCS approximation. 
Beyond mean-field calculations have already been carried out to study the evolution
of the low-lying excitation energies and transition strengths in Cd 
isotopes \cite{rodriguez_08} . The formalism used there involves a mean-field 
calculation with the finite range Gogny interaction, including particle number 
and angular momentum projections, as well as configuration mixing in the generator 
coordinate method. A different approach has also been explored in Ref. \cite{nomura_18}, 
where an IBM Hamiltonian was constructed to study the spectroscopy of Cd 
isotopes, based on a mapping of the triaxial energy surfaces 
obtained from constrained Skyrme self-consistent mean-field calculations. 
Similar calculations \cite{nomura_11,nomura2} in other mass regions have shown the
possibilities of that method.
A full understanding of the spectroscopic properties of Cd isotopes will require
in one way or another going beyond the mean-field approach. However, the question
to answer in this paper is to what extent a simple approach based on a self-consistent
mean field with standard Skyrme forces can reproduce the main features
observed experimentally on charge radii and quadrupole moments of ground and
isomeric states in Cd isotopes. 

The paper is organized as follows:
In the next section, the ingredients of the theoretical formalism used to calculate 
the nuclear properties are presented.
Section III  and IV contain the results obtained for quadrupole moments 
and charge radii, respectively. Section V contains the conclusions of the work.

\section{Theoretical approach}

The theoretical formalism used in this work is based on a self-consistent deformed 
Hartree-Fock (HF) mean-field calculation with effective two-body density-dependent 
Skyrme interactions, including pairing correlations in the BCS approximation.
Single-particle energies, wave functions, and occupation probabilities are 
generated from this mean field.
The virtues of this type of self-consistent microscopic formalisms are well known.
They include the universality of the Skyrme density functional that allows one
to use the same interaction throughout the entire nuclear chart. The reliability 
of such interactions also implies a great predictive power to those calculations.

The Skyrme interaction SLy4 \cite{chabanat} is used in this work as a representative 
reference of the Skyrme forces.
For comparison, I also consider the 
parametrization SkM* \cite{bartel}, which is also a reliable standard reference 
for Skyrme forces.
Actually, one can find in the literature hundreds of Skyrme interactions,
see for example Ref. \cite{dutra_12}, where 240 different parametrizations available 
at that time were considered. 
No doubt, other more sophisticated Skyrme interactions could be used for the
present study than those used here, but most of these 
new constructed interactions have been especially designed and adjusted to treat 
specific nuclear properties. Therefore, in this work I preferred to check the 
properties of Cd isotopes with standard and classical Skyrme forces with a
proven wide range of applicability that have been successfully tested throughout
the nuclear chart for a large variety of different nuclear properties \cite{bender}.
This is the case of the forces considered SkM* and SLy4. SkM* is based 
on an accurate description of nuclear ground states properties like radii and 
multipole moments, as well as surface energies and fission barriers. On the 
other hand, SLy4 was developed to describe properly neutron-rich nuclei.
These two forces are still meaningful references with which new developed 
Skyrme interactions are compared.

The solution of the HF equation, assuming time reversal and axial symmetry, is found 
by using the formalism developed in Ref. \cite{vautherin}. The single-particle wave 
functions are expanded in terms of the eigenstates of an axially symmetric harmonic 
oscillator in cylindrical coordinates, using twelve major shells. The two parameters 
of the cylindrical basis related to the oscillator length and the ratio of the axes 
are chosen optimally to minimize the energy. 
It is also worth noting that, contrary to the shell model approach where a restricted 
valence space is used, in this approach a large basis space is used and therefore,
core polarization effects are fully taken into account without effective charges.

The method also includes pairing between like nucleons in the BCS approximation with 
fixed gap parameters for protons and neutrons \cite{vautherin}, which are determined 
phenomenologically from the odd-even mass differences through a symmetric five-term
formula involving the experimental binding energies when available 
\cite{exp_masses}. 
In those cases where experimental information for masses is still not available, 
the same pairing gaps as the closer isotopes measured are used. Given the 
importance that pairing correlations have in the microscopic explanation of both 
quadrupole moments \cite{zhao_14} and charge radii \cite{hammen_18} in the Cd 
isotopes, I also show for comparison in some instances, the results obtained 
with a fixed pairing strength $G_{p,n}$, parametrized with a standard dependence 
on $Z$ and $N$. The BCS equations are solved at the end of each HF iteration, 
generating occupation probabilities of the single-particle states that are 
included in the next HF iteration. Thus, the HF+BCS problem is solved 
self-consistently.

In a further step, constrained HF calculations are performed with a quadratic 
constraint \cite{flocard}, minimizing the energy under the restriction of 
keeping the nuclear deformation fixed. The resulting plots of the total energy versus 
deformation give the landscapes of nuclear stability and are called 
deformation-energy curves (DECs).

%%%%%%%%%%%%%%%%%%%%%%%%%%% Fig 1 %%%%%%%%%%%%%%%%%%%%%%%%%%%%%%%%%%%%%%%%%%%%%%%%%%%%%%%%
\begin{figure}[htb]
\centering
\includegraphics[width=80mm]{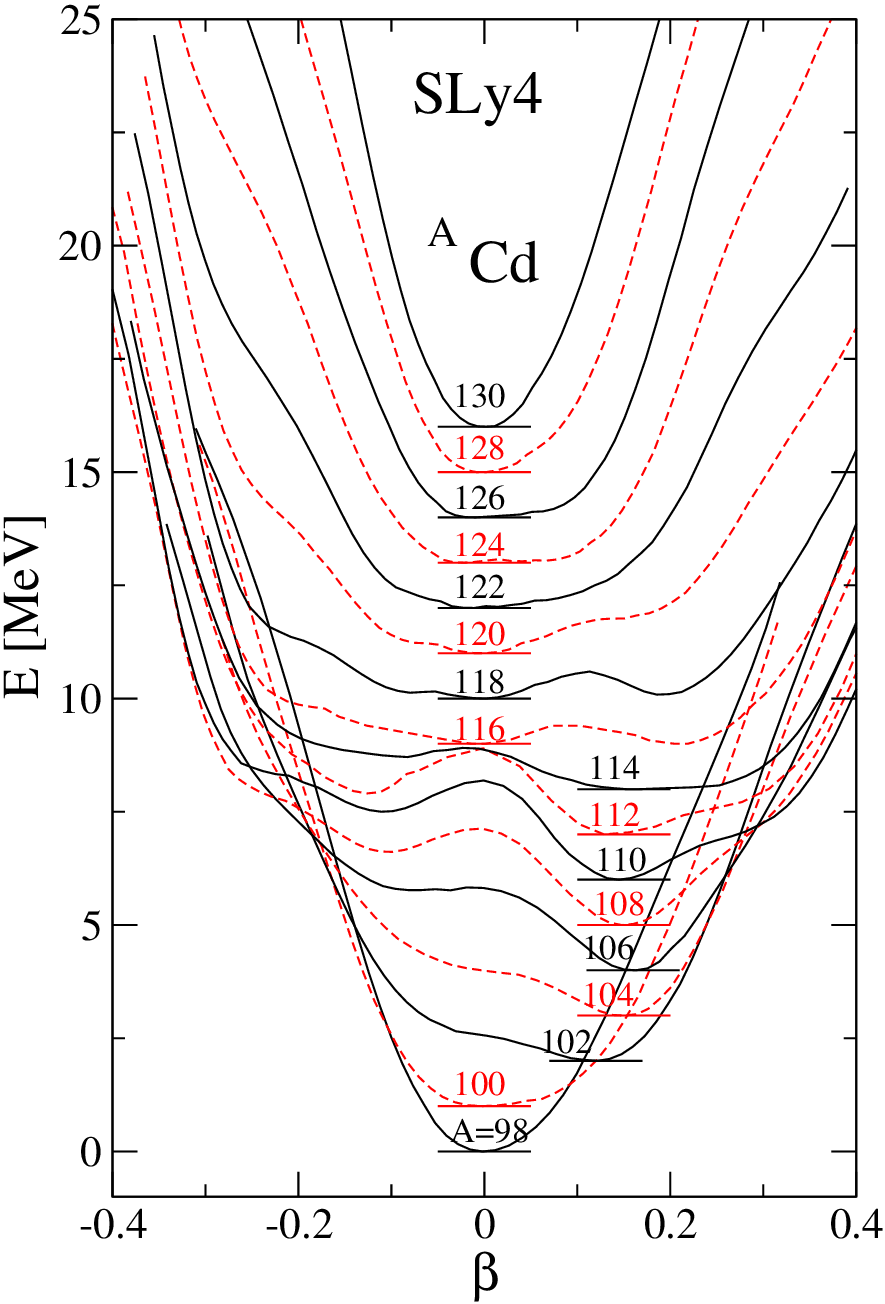}
\caption{Deformation-energy curves for even-even $^{98-130}$Cd isotopes obtained from 
constrained HF+BCS calculations with the Skyrme force SLy4. 
}
\label{fig_eq}
\end{figure}
%%%%%%%%%%%%%%%%%%%%%%%%%%%%%%%%%%%%%%%%%%%%%%%%%%%%%%%%%%%%%%%%%%%%%%%%%%%%%%%%%%%%%%%%%

Figure \ref{fig_eq} shows the DECs for even-even $^{98-130}$Cd isotopes that
are obtained from constrained HF+BCS calculations with the Skyrme force SLy4
and pairing correlations treated in the fixed-gap approach. The results are shown
as a function of the quadrupole deformation parameter $\beta$ calculated as
$\beta=\sqrt{\pi/5}Q_0/(A\langle r^2\rangle )$, which is defined in terms of 
the intrinsic quadrupole moment $Q_0$ and the nuclear mean-square radius 
$\langle r^2\rangle $ defined later. The DECs have
been scaled to the energy of their ground states and have been shifted by
1 MeV from one isotope to the next one, starting from the lightest one.

The isotopes studied start and end at neutron magic numbers, $A=98\ (N=50)$ and 
$A=130\ (N=82)$, respectively. It is observed that the spherical shapes at shell 
closures evolve to deformed shapes in the isotopes between them. First, as 
the number of neutrons increases, a prolate shape is developed with quadrupole 
deformations between $\beta=0.1$ and $\beta=0.2$. Around middle shell, between
$A=116$ and $A=126$, the DECs are rather shallow with deformations 
in the range of $-0.2 < \beta < 0.2$ that result practically degenerate 
in energy. Finally, spherical shapes are recovered as the shell closure at $A=130$
is approached. Similar trends in the DECs are obtained with other Skyrme forces, 
in particular, with the force SkM*. Thus, spherical, prolate, and oblate 
configurations can be identified in the profiles of the energy curves of mid-shell 
Cd isotopes. However, these profiles are very soft with fairly
shallow minima and shape coexistence could hardly be invoked. Rather they 
look like soft nuclei in this approach.

It is worth mentioning some existing works in this mass region based on mean-field 
approaches other than the present Skyrme HF + BCS calculations. In particular, 
mean-field studies of structural changes with the Gogny D1S interaction including
triaxiality have been systematically carried out \cite{gogny}.
Triaxial landscapes have been also studied in Ref. \cite{nomura_18} within Skyrme
mean-field calculations and the interacting boson model. These calculations 
show that the axial prolate and oblate minima are quite soft in both axial 
and triaxial directions. Indeed, the oblate axial minima become saddle 
points when the $\gamma$ degree of freedom is 
included in the analysis. The differences found in the axial equilibrium values
between the present HF + BCS approach and those works are not significant.

In the case of odd-$A$ nuclei, one-quasiparticle states are constructed by
allowing the unpaired nucleon to occupy a single-particle state, which is
blocked in the BCS calculation. The ground state is determined by finding 
the blocked state that minimizes the total energy.
Quasiparticle excitations correspond to configurations
with the odd nucleon in an excited state.
In the present study the equal filling approximation is used, a prescription
widely used in mean-field calculations to treat the dynamics of odd nuclei 
preserving time-reversal invariance \cite{rayner1}. In this approximation the 
unpaired nucleon is treated on an equal footing with its time-reversed state by 
sitting half a nucleon in a given orbital and the other half in the time-reversed 
partner. This approximation has been found \cite{schunck} to be equivalent to
the exact blocking when the time-odd fields of the energy density functional are 
neglected and therefore, it is sufficiently precise for most practical applications. 

Because the spin and parity of the ground and low-lying excited states of the 
nuclei studied are known experimentally,  the states to be blocked
are chosen according to those assignments.
In all cases the observed states 
have corresponding calculated states lying in the vicinity of the Fermi energy.

%%%%%%%%%%%%%%%%%%%%%%%%%%% Fig 2 %%%%%%%%%%%%%%%%%%%%%%%%%%%%%%%%%%%%%%%%%%%%%%%%%%%%%%%%
\begin{figure}[htb]
\centering
\includegraphics[width=80mm]{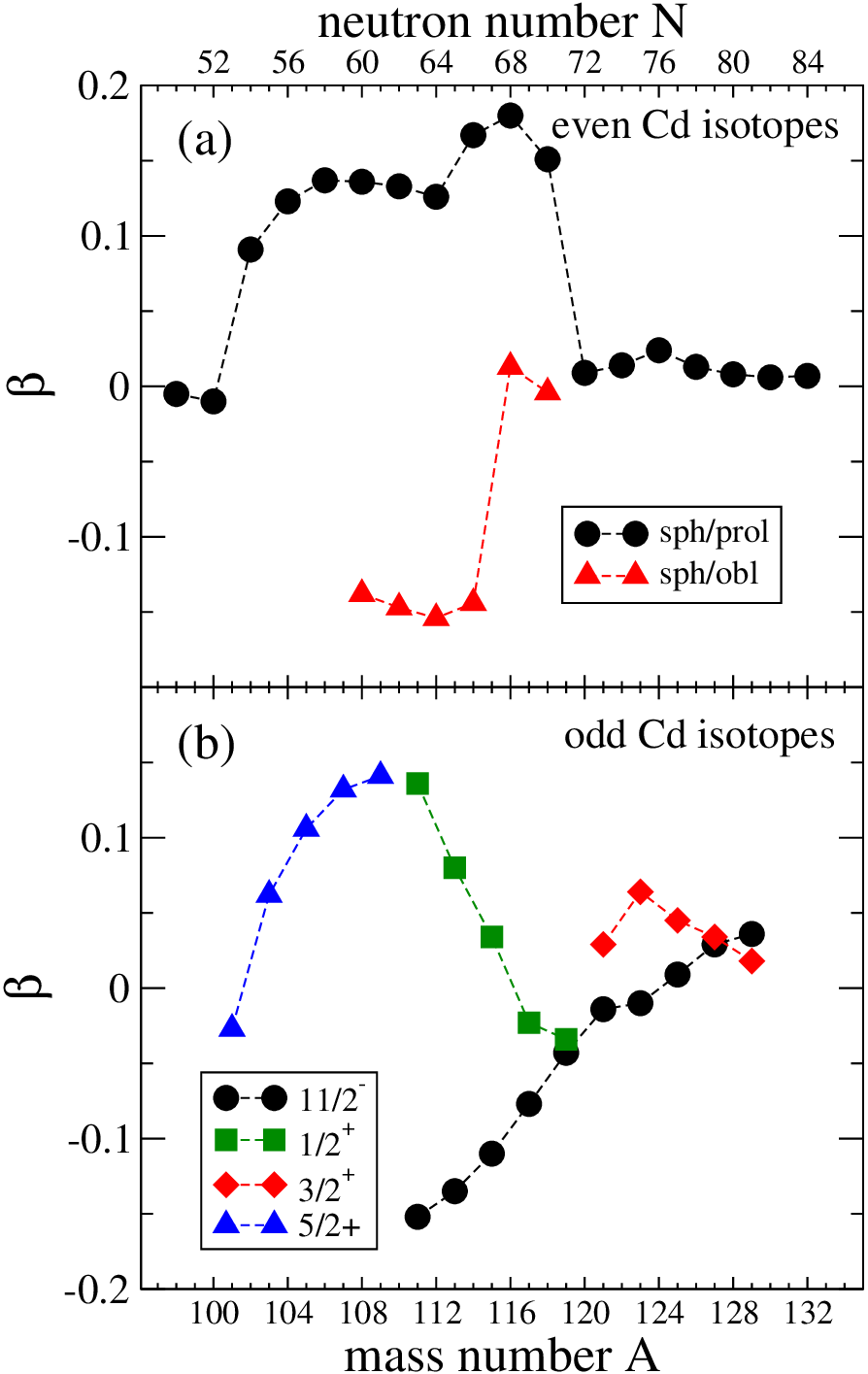}
\caption{Isotopic evolution of the quadrupole deformation $\beta$ in even-even (a) and
odd-$A$ nuclei (b) evaluated with the SLy4 interaction.
}
\label{fig_beta}
\end{figure}
%%%%%%%%%%%%%%%%%%%%%%%%%%%%%%%%%%%%%%%%%%%%%%%%%%%%%%%%%%%%%%%%%%%%%%%%%%%%%%%%%%%%%%%%%

%%%%%%%%%%%%%%%%%%%%%%%%%%% Fig 3 %%%%%%%%%%%%%%%%%%%%%%%%%%%%%%%%%%%%%%%%%%%%%%%%%%%%%%%%
\begin{figure}[t]
\centering
\includegraphics[width=80mm]{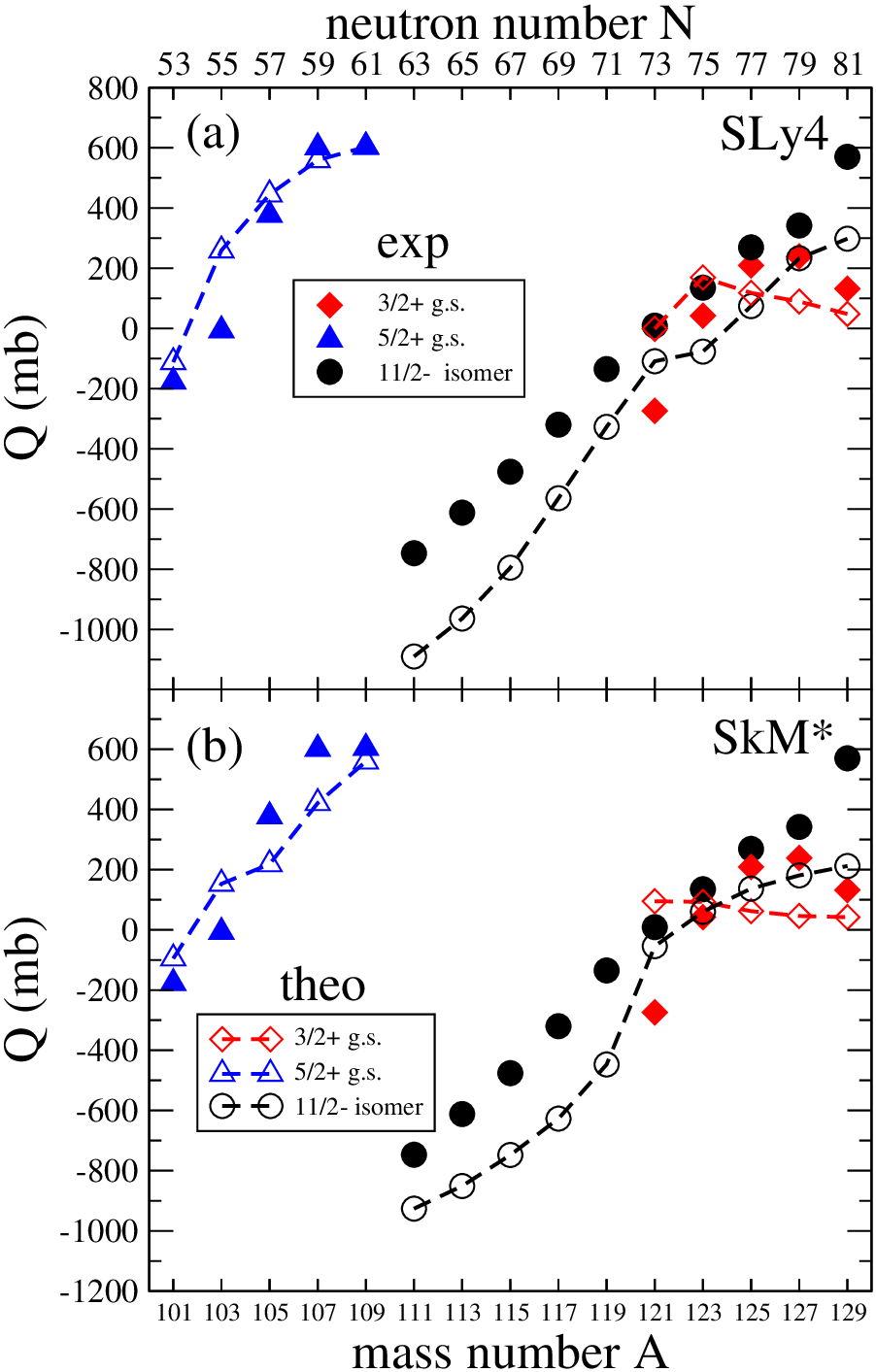}
\caption{Experimental quadrupole moments for the ground ($3/2^+,5/2^+$) and isomeric 
($11/2^-$) states \cite{yordanov_13,yordanov_18}, compared with HF+BCS calculations
with SLy4 (a) and SkM* (b) interactions.
}
\label{fig_q}
\end{figure}
%%%%%%%%%%%%%%%%%%%%%%%%%%%%%%%%%%%%%%%%%%%%%%%%%%%%%%%%%%%%%%%%%%%%%%%%%%%%%%%%%%%%%%%%%

\section{Quadrupole moments}

Figure \ref{fig_beta} (a) shows the isotopic evolution in the even-even Cd isotopes
of the quadrupole deformations $\beta$ associated with the various minima shown in 
Fig. \ref{fig_eq}. Also shown in panel (b) are the quadrupole deformations of the 
odd-$A$ isotopes. In this case the results for the ground states $5/2^+$ in 
$^{101-109}$Cd,  $1/2^+$ in $^{111-119}$Cd, and $3/2^+$ in $^{121-129}$Cd, as well 
as the deformations of the $11/2^-$ isomer states in $^{111-129}$Cd are shown.
From this figure, the progress of deformations with the number of nucleons, which
is already noticed in Fig. \ref{fig_eq}, is more apparent. Deformation increases 
as one departs from $N=50$, developing prolate and oblate structures with 
$\beta \approx \pm 0.15$. These deformed configurations become again spherical when 
approaching the next shell closure at $N=82$.
These results agree with recent large-scale shell-model calculations in the light 
Cd isotopes \cite{schmidt_17}.

In Fig. \ref{fig_q}, instead of the quadrupole deformations $\beta$ in the intrinsic 
system, the quadrupole moments in the laboratory frame are shown,

\begin{equation}
Q_{\rm lab}=\frac{3K^2-I(I+1)}{(I+1)(2I+3)} Q_0 \ ,
\end{equation}
where $Q_0$ is the intrinsic quadrupole moment of the proton distribution, 
\begin{equation}
Q_{0}=\sqrt{\frac{16\pi }{5}} \int \rho_p (\vec{r}) r^2 Y_{20}(\Omega) d\vec{r} \ ,
\end{equation}
$\rho_p (\vec{r})$ is the density of protons in the nucleus, $I$ is the total 
angular momentum and $K$ stands for its projection  along the symmetry axis.
The quadrupole moments are calculated within the HF+BCS approach with SLy4 (a) 
and SkM* (b) interactions. They are compared 
with the experimental values for the ground ($3/2^+,5/2^+$) and isomeric ($11/2^-$) 
states \cite{yordanov_13,yordanov_18}. 
Note that the quadrupole moments of the $1/2^+$ states vanish in the laboratory frame.

The isotopic linear increasing observed in the quadrupole moments of the $11/2^-$ 
isomeric states  was interpreted in terms of a basic shell model in 
Ref. \cite{yordanov_13}, splitting the total quadrupole moments into 
single-particle and core contributions. The data were reproduced with a simple fit  
$Q=((120-A)/9)Q_{\rm sp}+Q_{\rm core}$, where $Q_{\rm sp}=-667$ mb and $Q_{\rm core}=-85$ mb. 
The fitted value of $Q_{\rm sp}$ turns out to be about a factor of two larger than the 
expected value obtained from a neutron in the $h_{11/2}$ orbital, suggesting a 
strong polarization of the proton distribution. 

Microscopic calculations based on a relativistic mean-field approach \cite{zhao_14}
demonstrated the relevant role played by pairing correlations that soften the
stepped variations produced by the single-particle shell structure. It was also 
found that the core is strongly coupled with the valence nucleons, giving rise 
to important core polarization contributions to the quadrupole moments.

The general trend of the isotopic evolution observed experimentally is reproduced in 
the present calculations. In particular, the linear increase
of the quadrupole moments of the isomer $11/2^-$ states with $N$ is obtained,
although the calculations underestimate somewhat the data.
In the deformed formalism, the fact that the nuclear shape of the $11/2^-$ isomers
evolves from oblate to prolate with increasing neutron number has a simple explanation
in terms of Nilsson-like diagrams, where one plots the single-particle energies as
a function of the quadrupole deformation. The Nilsson state with asymptotic quantum 
numbers $[N\ n_z\ m_\ell]K=[505]11/2$ that corresponds to our $K^\pi =11/2^-$ states,
goes down (up) in energy with increasing deformation in the oblate (prolate) sector. 
Thus, the odd neutron in light isotopes will try to occupy that state in the oblate 
sector, which is lower in energy. In heavier isotopes with more neutrons occupying 
lower energy states, the state $[505]11/2$ will be occupied in the prolate sector.

%%%%%%%%%%%%%%%%%%%%%%%%%%% Fig 4 %%%%%%%%%%%%%%%%%%%%%%%%%%%%%%%%%%%%%%%%%%%%%%%%%%%%%%%%
\begin{figure}[t]
\centering
\includegraphics[width=80mm]{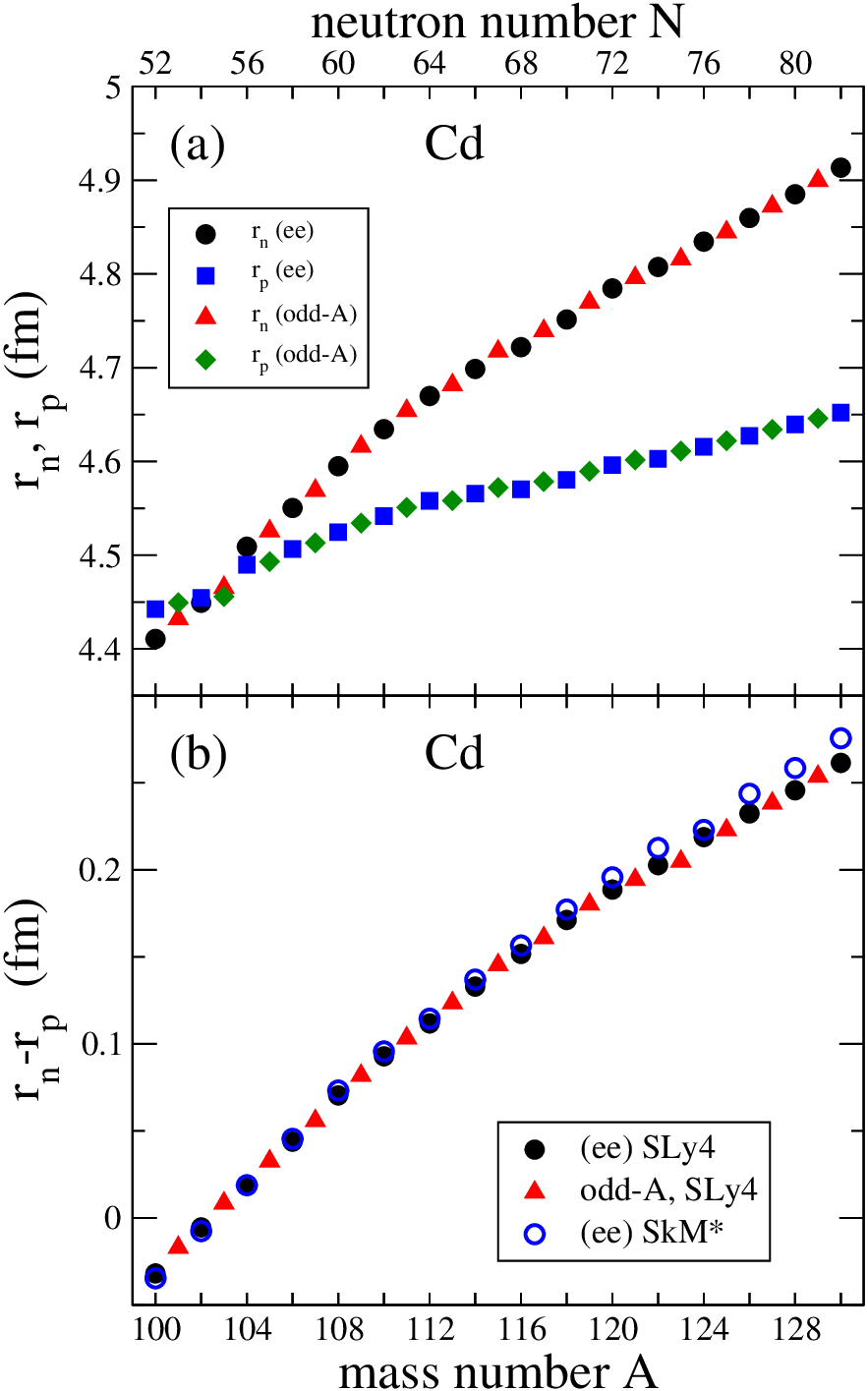}
\caption{(a) Neutron and proton root-mean-square radii calculated with SLy4 HF+BCS. 
(b) Difference $r_n-r_p$ calculated with SLy4 and SkM*.
}
\label{fig_rnp}
\end{figure}
%%%%%%%%%%%%%%%%%%%%%%%%%%%%%%%%%%%%%%%%%%%%%%%%%%%%%%%%%%%%%%%%%%%%%%%%%%%%%%%%%%%%%%%%%

%%%%%%%%%%%%%%%%%%%%%%%%%%% Fig 5 %%%%%%%%%%%%%%%%%%%%%%%%%%%%%%%%%%%%%%%%%%%%%%%%%%%%%%%%
\begin{figure}[t]
\centering
\includegraphics[width=80mm]{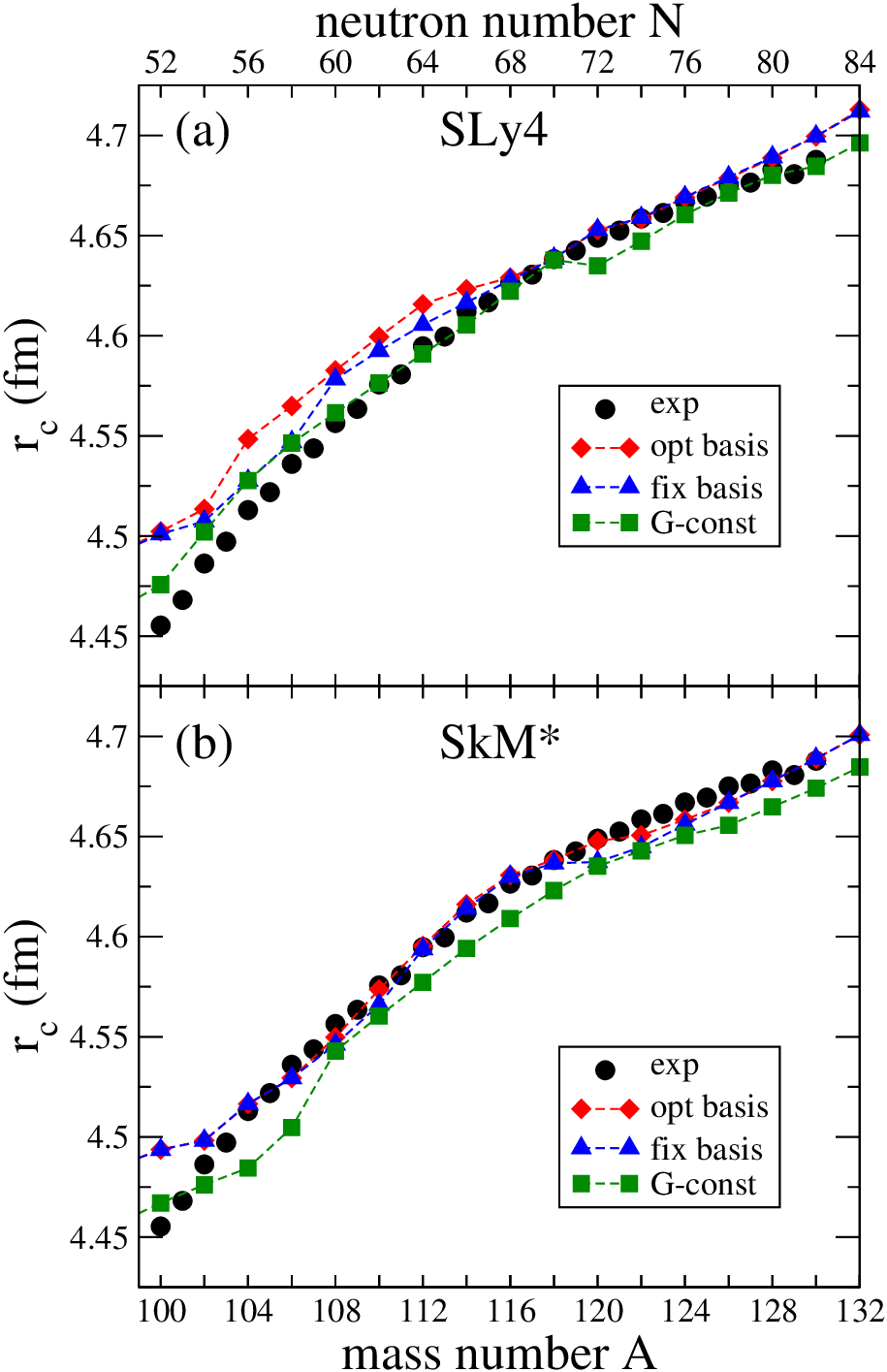}
\caption{Root-mean-square charge radii of the ground states in $^{100-132}$Cd isotopes.
Experimental data \cite{hammen_18} are compared with calculations from SLy4 (a) and 
SkM* (b) with various theoretical treatments 
of the pairing and basis parameters (see text).
}
\label{fig_rc}
\end{figure}
%%%%%%%%%%%%%%%%%%%%%%%%%%%%%%%%%%%%%%%%%%%%%%%%%%%%%%%%%%%%%%%%%%%%%%%%%%%%%%%%%%%%%%%%%

%%%%%%%%%%%%%%%%%%%%%%%%%%% Fig 6 %%%%%%%%%%%%%%%%%%%%%%%%%%%%%%%%%%%%%%%%%%%%%%%%%%%%%%%%
\begin{figure}[t]
\centering
\includegraphics[width=80mm]{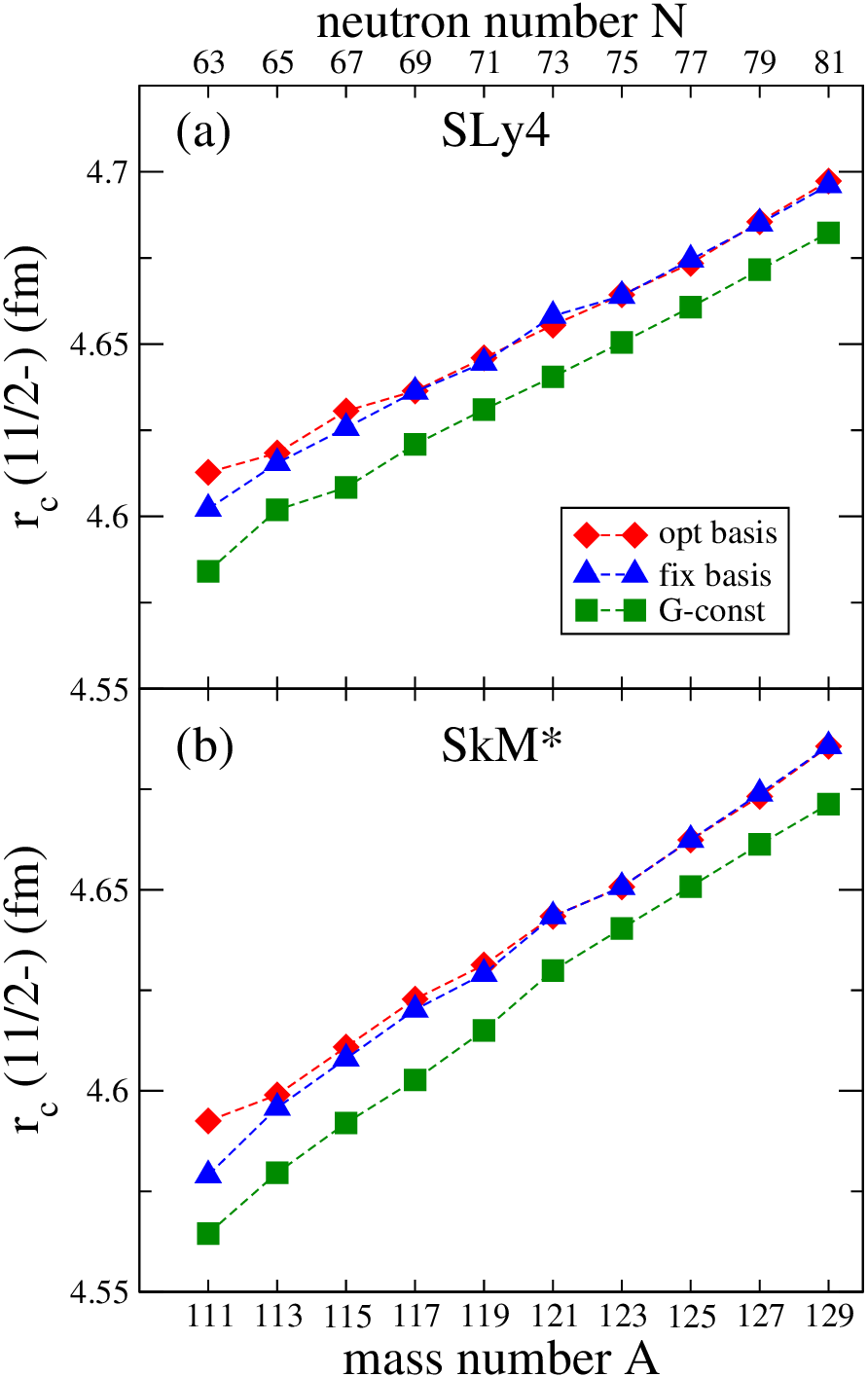}
\caption{Root-mean-square charge radii of the isomer $11/2^-$ states in $^{111-129}$Cd 
odd-$A$ isotopes. Calculations are from SLy4 (a) and 
SkM* (b) with various theoretical approaches.}
\label{fig_rc_112}
\end{figure}
%%%%%%%%%%%%%%%%%%%%%%%%%%%%%%%%%%%%%%%%%%%%%%%%%%%%%%%%%%%%%%%%%%%%%%%%%%%%%%%%%%%%%%%%%

%%%%%%%%%%%%%%%%%%%%%%%%%%% Fig 7 %%%%%%%%%%%%%%%%%%%%%%%%%%%%%%%%%%%%%%%%%%%%%%%%%%%%%%%
\begin{figure}[t]
\centering
\includegraphics[width=80mm]{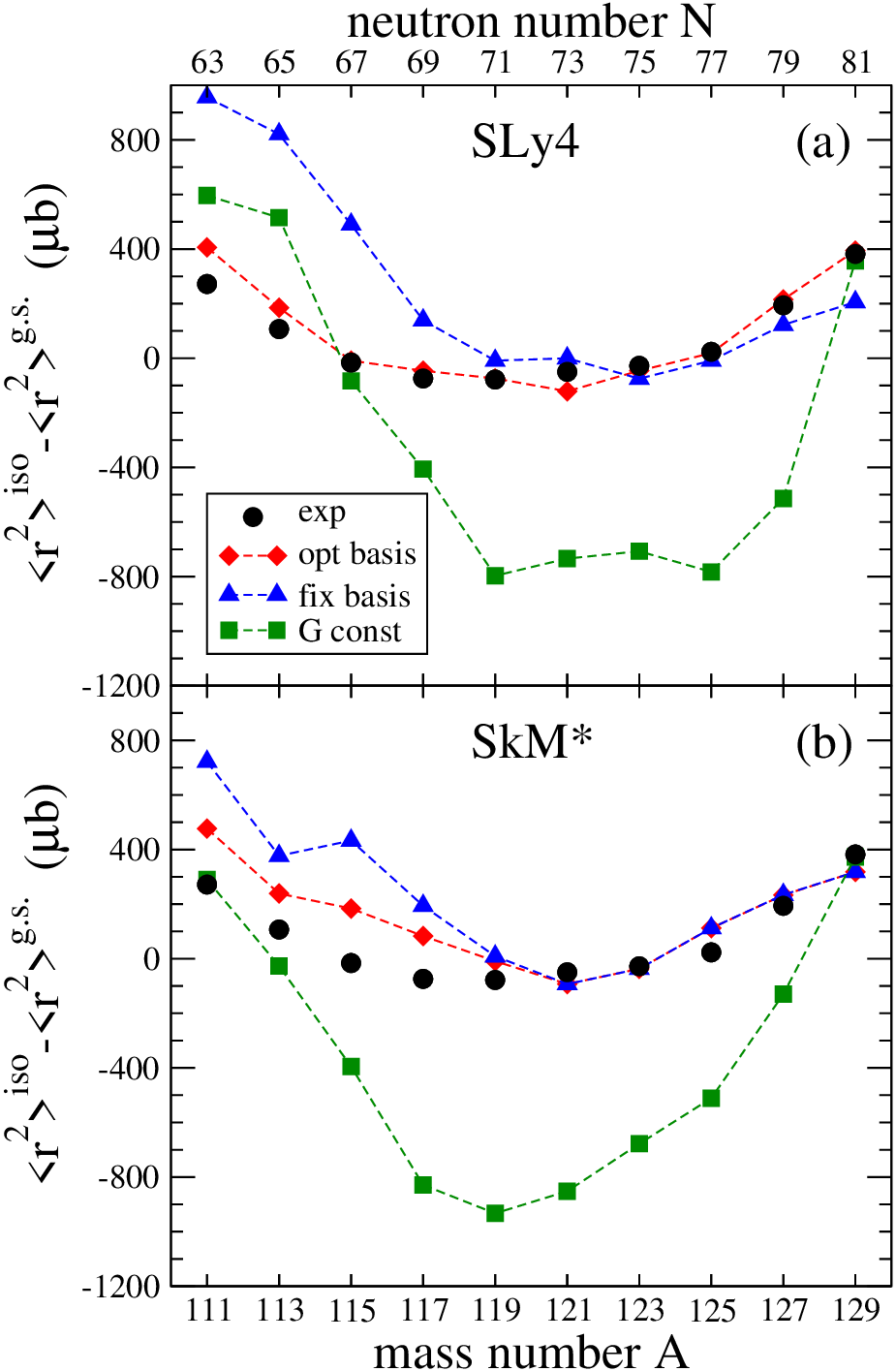}
\caption{Experimental mean-square charge radii differences between $11/2^-$ isomers and 
$1/2^+,3/2^+$ ground states \cite{yordanov_16} 
compared with Skyrme HF+BCS calculations with SLy4 (a) and SkM* (b). 
}
\label{fig_delta_r_isom}
\end{figure}
%%%%%%%%%%%%%%%%%%%%%%%%%%%%%%%%%%%%%%%%%%%%%%%%%%%%%%%%%%%%%%%%%%%%%%%%%%%%%%%%%%%%%%%%%

\section{Charge radii}

Charge radii and their isotopic differences  have been shown 
\cite{rayner2,rayner3,boillos,sarri_15}
to be suitable quantities to study the nuclear shape evolution as they can be measured 
with 
remarkable precision using laser spectroscopic techniques \cite{cheal,campbell}. 
One of the most noticeable characteristics of the isotopic evolution of charge radii 
is probably the kink that appears at shell closures. 
Sudden changes in the charge radii behavior are also 
good indicators of nuclear shape transitions taking place when
the addition of a single neutron induces collective nuclear effects that drives
the nucleus from a spherical into a deformed shape or vice versa. 
The odd-even staggering in certain neighbor isotopes is another manifestation 
related to the blocking effect of the odd nucleon.

There are in the literature a large variety of theoretical calculations aiming to
account for the nuclear charge radii.
The different approaches include phenomenological effective formulas
that go beyond the simplest $r_0A^{1/3}$ expression to various degrees of sophistication
\cite{nerlo,sheng_15};  Garvey-Kelson relations \cite{piekarewicz_10};
macroscopic-microscopic models \cite{moeller,buchinger,iimura_08,iimura_15}; and
fully microscopic models either relativistic \cite{lalazissis,long,geng_05}
or non-relativistic with Gogny \cite{gogny,libert,delaroche} and Skyrme forces
\cite{bender,goriely}.

The nuclear  mean-square proton and neutron radii are defined as
\begin{equation}
\langle r_{\rm p,n}^2 \rangle =\frac{ \int r^2\rho_{\rm p,n}({\vec r})d{\vec r}}
{\int \rho_{\rm p,n}({\vec r})d{\vec r}} \, , \label{r2pn}
\end{equation}
and the corresponding root-mean-square radii for protons and neutrons are simply 
given by
\begin{equation}
r_{\rm p,n}= \langle r_{\rm p,n}^2 \rangle ^{1/2} \, . \label{rmsrnp}
\end{equation}

Figure \ref{fig_rnp} shows in the upper panel (a), the neutron and proton 
root-mean-square 
radii for both even and odd Cd isotopes. Proton radii are larger than
neutron radii up to $N=54$ because of the Coulomb repulsion of the protons. For
heavier isotopes the neutron radii increase more rapidly than the proton radii
that still increase following the neutrons in spite of the fixed number of
protons. Panel (b) shows the neutron skin thickness, $r_n-r_p$
that increases continuously
as the number of neutrons increases. The figure includes also the results with SkM*
that do not differ much from SLy4.
Microscopic Skyrme HF+BCS calculations similar to those presented here 
have been performed in various isotopic chains \cite{skin,gaidarov1,gaidarov2}.

The mean-square radius of the nuclear charge distribution in a nucleus is calculated 
by folding the proton distribution of the nucleus with the finite size of the protons 
and the neutrons. It can be expressed as \cite{negele,bertozzi}
\begin{equation}
\langle r^2_c \rangle = \langle r^2_p \rangle _Z+
\langle r^2_c \rangle _p +(N/Z)
\langle r^2_c \rangle _n + r^2_{CM} + r^2_{\rm SO}
\, , \label{rch1}
\end{equation}
where $ \langle r^2_p \rangle _Z$ is the mean-square radius of the point-proton 
distribution in the nucleus given by Eq. (\ref{r2pn}).
$ \langle r^2_c \rangle _p=0.80$ fm$^2$ \cite{sick03} and 
$ \langle r^2_c \rangle _n=-0.12$ fm$^2$ \cite{gentile11} are the mean-square radii 
of the charge distributions in a proton and a neutron, respectively. 
$r^2_{CM}$ is a small correction due to the center of mass motion,
which is evaluated assuming harmonic-oscillator wave functions. The last term 
$r^2_{\rm SO}$ is also a small spin-orbit contribution to the charge density. 
The root-mean-square charge radius of the nucleus, $r_{\rm c}$, is simply given by
\begin{equation}
r_{\rm c}= \langle r_{\rm c}^2 \rangle  ^{1/2} \, . \label{rmsrc}
\end{equation}

It is worth noting that the most important correction to the point proton 
mean-square nuclear radius, coming from the charge distribution of the proton 
$ \langle r^2_c \rangle _p$, vanishes when isotopic differences $r^2(Z,A)-r^2(Z,A')$ 
are considered, since it does not involve any dependence on $N$.
Nuclear structure effects are also canceled to a large extent in isotopic differences.
In the case of radii differences between isomers and ground states 
$ r^2(Z,A)_{\rm isomer} - r^2(Z,A)_{\rm gs}$, all the corrections due to finite sizes
of nucleons and center of mass are canceled in the differences and only the
point proton mean-square radii $\langle r^2_p \rangle _Z$ of the isomer and ground 
state are involved.
These radii depend only on the density of protons in the nucleus, which in turn
reflects how a given number of them ($Z=48$) are redistributed to accommodate 
the changing neutron environment.

Figure \ref{fig_rc} shows the absolute values of the charge radii $r_c$, as measured
in Ref. \cite{hammen_18}, taking $r_c(^{114}$Cd)=4.612 fm as the reference radius.
They are compared with the Skyrme HF+BCS results using SLy4 (a) and SkM* (b).
Various theoretical treatments for the pairing interaction and different parameters 
of the cylindrical basis used in the expansion of the HF wave functions are considered. 
First, the effect of the 
basis is studied by comparing the results obtained with a fixed basis for all the 
isotopes (spherical basis with a given oscillator length) with the results obtained 
from an optimal choice of the deformed basis singularized for each isotope in order
to minimize its energy. These results are labeled 'fix basis' and 'opt basis', 
respectively. Two different treatments of the pairing correlations are also considered
by solving the BCS equations with either fixed gap parameters,
fitted phenomenologically to the experimental masses, or with a fixed strength of 
the pairing force, labeled 'G const' in the figure. 

The general trend observed experimentally is reproduced by the
calculations. The results with constant $G$-pairing appear systematically 
below the results with constant pairing gaps, whereas the difference  in
the radii calculated with different bases is more sizable in the lighter 
isotopes because in this region nuclei are more deformed and therefore, the role 
of the deformed  basis is expected to be more
critical in the description of the equilibrium configurations.

Similarly, Fig. \ref{fig_rc_112} shows the root-mean-square charge radii of the 
isomer $11/2^-$ states in the $^{111-129}$Cd odd-$A$ isotopes. Calculations are also
from SLy4 (a) and  SkM* (b) with the same theoretical approaches discussed in
the previous figure for the ground-state radii.
The behavior observed from different treatments of pairing and basis is
similar to that in the previous figure. That is, very similar radii with the
two options for the basis, except in the lighter and deformed isotopes,
and a shift to lower radii when $G$-paring is considered.

As mentioned above,
the variations of the charge radii patterns in isotopic chains can be related
to deformation effects that can be used as signatures of shape transitions.
This can be understood easily by expressing the radius as a function of
deformation in a simple liquid drop model.
For a nucleus with an axially symmetric static quadrupole deformation $\beta$ 
the increase of the charge radius with respect to the radius of the spherical 
nucleus is given to first order by

\begin{equation}
\langle r^2 \rangle = \langle r^2 \rangle _{\rm sph} \left(
1+\frac{5}{4\pi} \beta^2 \right) \, ,
\label{rbeta}
\end{equation}
where  $\langle r^2 \rangle _{\rm sph}$ is the mean-square radius of a spherical nucleus
with the same volume, usually taken from the droplet 
model. In this work I analyze the effect of the quadrupole deformation on 
the charge radii from a microscopic self-consistent approach.
The fact that there is a smooth evolution with no apparent jumps in Figs. \ref{fig_rc} 
and \ref{fig_rc_112} tells us that no sudden shape transition in the 
evolution of Cd isotopes is expected.

Mean-square charge radii differences between isomers and ground states in odd-$A$ 
Cd isotopes
were measured in Ref. \cite{yordanov_16}. A parabolic pattern of these differences
was found as a function of the neutron number that was explained on the basis of
a simple model as a consequence of the combination of the linear increase of the 
quadrupole moment of the $11/2^-$ isomer states and the rather constant behavior of 
the ground state quadrupole moments. The minimum of the parabola corresponds 
to the zero of the quadrupole moment of the isomer. It was also found that the
results from a relativistic mean-field approximation support those 
findings \cite{yordanov_16}.

The results for the charge radii differences between isomers and ground states
within our HF+BCS approach
are shown in Fig. \ref{fig_delta_r_isom} for SLy4 (a) and SkM* (b).
The figure also explores the sensitivity of the results to the pairing force
and to the parameters of the deformed basis.
In general, the results with the optimal basis describe fairly well the observed 
parabolic behavior. The fixed basis reproduces also quite well the heavier isotopes,
but becomes somewhat worse in lighter ones. As explained above, the reason for this 
is related to the larger deformation of lighter isotopes.
The results with the constant pairing interaction show also a parabolic pattern, 
but they fit much worse the measurements. The pairing interaction that treats the
gaps as phenomenological input parameters reproduces better the observed behavior.
Results from SkM* in panel (b) are qualitatively similar.

One should keep in mind that because of the high precision achieved in the 
measurements of the difference  between isomeric and ground state  mean-square 
charge radii, which is of the order of 10 $\mu$b  \cite{yordanov_16}, the 
theoretical accuracy reached in present calculations of charge radii is
not enough to predict unambiguously this behavior.
In the particular case of Cd isotopes, there is an additional difficulty
because the DECs (see Fig. \ref{fig_eq}) show rather shallow patterns
around the minima. Therefore, nuclear configurations with quite different 
deformations have practically the same energy, but according to Eq. (\ref{rbeta}),
they have different charge radii.
Thus, very little changes in the input parameters of the method have almost 
no effect on the energies (minimizing the energy is, for example, the criterion 
for choosing the optimal basis), but it prevents the radii from being determined 
with a high level of precision.

Actually, no present theory is able to provide the level of accuracy reached
experimentally. Typical mean deviations 
$\left( \bar{\epsilon}=\langle r_c({\rm theo})-r_c({\rm exp})\rangle \right) $ 
and root-mean-square deviations
$\left( \sigma=\langle \left[ r_c({\rm theo})-r_c({\rm exp})\right] ^2
\rangle ^{1/2}\right)$ of different theoretical calculations depend of course of 
the sample size considered, but in general they have values around 
$\bar{\epsilon} = 0.001$ fm and $\sigma =0.026$ fm in mean-field calculations with 
modern Skyrme interactions \cite{goriely,goriely_16}. 
Mean deviations of around  $\bar{\epsilon}=0.03-0.04$ fm are also found with the 
D1S and D2 Gogny interactions \cite{delaroche,pillet}.
These uncertainties are comparable to those obtained with other more phenomenological
models, such as macroscopic-microscopic models  \cite{iimura_15}, where values of
$\sigma=0.04-0.08$ fm are found; models based on Garvey-Kelson relations 
\cite{piekarewicz_10} with $\sigma=0.01$ fm; and purely effective formulas
\cite{sheng_15}, where a value $\sigma=0.022$ fm is found.
Although these theoretical uncertainties are still quite large, they are
reduced to some extent when dealing with radii differences between isomers and 
ground states due to the cancellation of the effects of the common nuclear structure. 
The underlying parabolic behavior of these differences would be the signature 
of such cancellations. The only surviving effect would be the different
quadrupole deformations of ground and isomer states, leading to the observed 
parabolic behavior associated with the increasing deformations from oblate
to prolate in the isomers and the almost constant deformations
of the ground states.

\section{Conclusions}

This work presents results of quadrupole moments and charge radii in even and odd
$^{98-130}$Cd isotopes. Microscopic calculations based on self-consistent deformed 
Skyrme Hartree-Fock with pairing are compared with high resolution laser spectroscopy 
experiments.
In the case of odd-$A$ isotopes, both ground states and $11/2^-$ isomeric states
are studied.
The main characteristics observed in the isotopic evolution, such as a linear increase
of the quadrupole moments of the $11/2^-$ isomers with the number of neutrons and 
a parabolic behavior of the mean-square charge radii difference between 
isomers and ground states, are well accounted for by the calculations.

The results obtained with both Skyrme forces, SLy4 and SkM*, are qualitatively
similar with a reasonable agreement with the experiment.
This does not mean that other Skyrme 
interactions could reproduce the measurements better or that they could completely 
fail to do that. The results in this paper tell us that standard 
Skyrme forces, which are known to be quite robust to describe a large 
variety of nuclear properties in very different mass regions, are also 
suitable to explain the phenomenology observed in the charge radii behavior 
of Cd isotopes.

The rather shallow DECs obtained in the isotopes of cadmium make the analysis of 
nuclear radii especially difficult due to the uncertainty that arises from having 
different deformations and, therefore, different radii with practically the 
same binding energy.

Although theoretical uncertainties due to different methods and approximations,
or due to different interactions and parameter choices 
cannot compete with the high resolution achieved experimentally, 
still the main features observed experimentally are well described.
This is specially true in the case of radii differences, where the 
cancellation of nuclear structure effects common to ground and 
isomeric states, makes evident the underlying parabolic behavior.

\begin{acknowledgments}
This work was supported by Ministerio de Ciencia, Innovaci\'on y Universidades 
MCIU/AEI/FEDER,UE (Spain) under Contract No. PGC2018-093636-B-I00.  

\end{acknowledgments}


\begin{thebibliography}{99}

\bibitem{cheal} B. Cheal and K. T. Flanagan, J. Phys. G: Nucl. Part. Phys. 
{\bf 37}, 113101 (2010).

\bibitem{campbell} P. Campbell, I. D. Moore, and M. R. Pearson, Prog. Part. Nucl.
Phys. {\bf 86}, 127 (2016).

\bibitem{wood-heyde} J. L. Wood, K. Heyde, W. Nazarewics, M. Huyse, and P. van Duppen,
Phys. Rep. {\bf 215}, 101 (1992).

\bibitem{bender_03} M. Bender, P.-H. Heenen, and P.-H. Reinhard, Rev. Mod. Phys.
{\bf 75}, 131 (2003).

\bibitem{angeli} I. Angeli and  K. P. Marinova, At. Data Nucl. Data Tables 
{\bf 99}, 69 (2013).

\bibitem{stone} N. J. Stone, J. Phys. Chem. Ref. Data {\bf 44}, 031215 (2015).

\bibitem{garrett_10} P. E. Garrett and J. L  Wood, J. Phys. G: Nucl. Part. Phys. 
{\bf 37}, 064028 (2010); {\it Corrigendum} 069701.

\bibitem{heyde_11} K. Heyde and J. L. Wood, Rev. Mod. Phys.  {\bf 83}, 1467 (2011).

\bibitem{heyde_04} K. Heyde, J. Jolie, R. Fossion, S. De Baerdemacker, and V. Hellemans,
Phys. Rev. C {\bf 69}, 054304 (2004).

\bibitem{yordanov_13} D. T. Yordanov, D. L. Balabanski, J. Bieron, M. L. Bissell,
K. Blaum, I. Budincevic, S. Fritzsche, N. Fr\"ommgen, G. Georgiev, Ch. Geppert,
M. Hammen, M. Kowalska, K. Kreim, A. Krieger, R. Neugart, W. N\"ortersh\"auser,
J. Papuga, and S. Schmidt,
Phys. Rev. Lett.  {\bf 110}, 192501 (2013).

\bibitem{yordanov_16} D. T. Yordanov, D. L. Balabanski, M. L. Bissell, K. Blaum,
I. Budincevic, B. Cheal, K. Flanagan, N. Fr\"ommgen, G. Georgiev, Ch. Geppert,
M. Hammen, M. Kowalska, K. Kreim, A. Krieger, J. Meng, R. Neugart, G. Neyens,
W. N\"ortersh\"auser, M. M. Rajabali, J. Papuga, S. Schmidt, and P.W. Zhao,
Phys. Rev. Lett.  {\bf 116}, 032501 (2016).

\bibitem{yordanov_18} D. T. Yordanov, D. L. Balabanski, M. L. Bissell, K. Blaum,
A. Blazhev, I. Budincevic, N. Fr\"ommgen, Ch. Geppert, H. Grawe, M. Hammen,
K. Kreim, R. Neugart, G. Neyens, and W. N\"ortersh\"auser, 
Phys. Rev. C {\bf 98}, 011303(R) (2018).

\bibitem{hammen_18} M. Hammen, W. N\"ortersh\"auser, D. L. Balabanski, M. L. Bissell,
K. Blaum, I. Budincevic, B. Cheal, K. T. Flanagan, N. Fr\"ommgen, G. Georgiev,
Ch. Geppert, M. Kowalska, K. Kreim, A. Krieger, W. Nazarewicz, R. Neugart,
G. Neyens, J. Papuga, P.-G. Reinhard, M. M. Rajabali, S. Schmidt, and D. T. Yordanov,
Phys. Rev. Lett.  {\bf 121}, 102501 (2018).

\bibitem{zhao_14} P. W. Zhao, S. Q. Zhang, and J. Meng, 
Phys. Rev. C {\bf 89}, 011301(R) (2014).

\bibitem{reinhard_17} P.-G. Reinhard and W. Nazarewicz,  Phys. Rev. C {\bf 95} 
064328 (2017).

\bibitem{rodriguez_08} T. R. Rodr\'{\i}guez, J. L. Egido, and A. Jungclaus, 
Phys. Lett. B {\bf 668}, 410 (2008).

\bibitem{nomura_18} K. Nomura and J. Jolie, Phys. Rev. C {\bf 98} 024303 (2018).

\bibitem{nomura_11} K. Nomura, T. Otsuka, R. Rodr\'{\i}guez-Guzm\'an,  L. M. Robledo, 
and P. Sarriguren, Phys. Rev. C {\bf 83}, 014309 (2011).

\bibitem{nomura2}  K. Nomura, T. Otsuka, R. Rodr\'{\i}guez-Guzm\'an,  L. M. Robledo, 
P. Sarriguren, P. H. Regan, P. D. Stevenson, and Z. Podolyak,   
Phys. Rev. C {\bf 83}, 054303 (2011); {\it ibid} {\bf 84}, 054316 (2011).

\bibitem{chabanat} E. Chabanat, P. Bonche, P. Haensel, J. Meyer, and
R. Schaeffer, Nucl. Phys. A {\bf 635}, 231 (1998).

\bibitem{bartel} J. Bartel, P. Quentin, M. Brack, C. Guet, and H.-B. Hakansson,
Nucl. Phys. A {\bf 386}, 79 (1982).

\bibitem{dutra_12} M. Dutra, O. Louren\c co, J. S. S\'a Martins, A. Delfino, 
J. R. Stone, and P. D. Stevenson, Phys. Rev. C {\bf 85}, 035201 (2012).

\bibitem{bender} M. V. Stoitsov, J. Dobaczewski, W. Nazarewicz, S. Pittel, and
D. J. Dean, Phys. Rev. C {\bf 68}, 054312 (2003); \\
https://www.fuw.edu.pl/$\sim$dobaczew/thodri/thodri.html

\bibitem{vautherin} D. Vautherin and D. M. Brink, Phys. Rev. C
{\bf 5}, 626 (1972); D. Vautherin, Phys. Rev. C {\bf 7}, 296 (1973).

\bibitem{exp_masses} G. Audi, F. G. Kondev, M. Wang, B. Pfeiffer, X. Sun, J.
Blachot, and M. MacCormick, Chinese Phys. C {\bf 36}, 1157 (2012);
M. Wang, G. Audi, A. H. Wapstra, F. G. Kondev, M. MacCormick, X. Xu, and 
B. Pfeiffer, {\it ibid.} {\bf 36}, 1603 (2012).

\bibitem{flocard} H. Flocard, P. Quentin, A. K. Kerman, and D. Vautherin, Nucl.
Phys. A {\bf 203}, 433 (1973).

\bibitem{gogny} S. Hilaire and M. Girod, Eur. Phys. J. A {\bf 33}, 237 (2007);\\ 
 www-phynu.cea.fr/science\_en\_ligne/carte\_potentiels\-
\_microscopiques/carte\_potentiel\_nucleaire\_eng.htm

\bibitem{rayner1} R. Rodr\'{\i}guez-Guzm\'an, P. Sarriguren, and L. M. Robledo,
 Phys. Rev. C {\bf 82}, 044318 (2010). 

\bibitem{schunck}  N. Schunck, J. Dobaczewski, J. McDonnell, J. Mor\'e, W.
Nazarewicz, J. Sarich, and M. V. Stoitsov, 
Phys. Rev. C {\bf 81}, 024316 (2010).

\bibitem{schmidt_17} T. Schmidt, K. L. G. Heyde, A. Blazhev, and J. Jolie, 
Phys. Rev. C {\bf 96}, 014302 (2017).

\bibitem{rayner2} R. Rodr\'{\i}guez-Guzm\'an, P. Sarriguren, L. M. Robledo, and 
S. Perez-Martin,   Phys. Lett. B {\bf 691}, 202 (2010).

\bibitem{rayner3} R. Rodr\'{\i}guez-Guzm\'an, P. Sarriguren, and L. M. Robledo
Phys. Rev. C {\bf 82}, 061302(R) (2010);
Phys. Rev. C {\bf 83}, 044307 (2011).

\bibitem{boillos} J. M. Boillos and P. Sarriguren, Phys. Rev. C {\bf 91}, 034311 (2015).

\bibitem{sarri_15} P. Sarriguren, Phys. Rev. C {\bf 91}, 044304 (2015).

\bibitem{nerlo} B. Nerlo-Pomorska and K. Pomorski, Z. Phys. A {\bf 348}, 169 (1994).

\bibitem{sheng_15} Zongqiang Sheng, Guangwei Fan, Jianfa Qian, and Jigang Hu, 
Eur. Phys. J. A {\bf 51}, 40 (2015).

\bibitem{piekarewicz_10} J. Piekarewicz, M. Centelles, X. Roca-Maza, and X. Vi\~nas,
Eur. Phys. J. A {\bf 46}, 379 (2010).

\bibitem{moeller} P. M\"oller, J. R. Nix, W. D. Myers, and W. J. Swiatecki, 
At. Data Nucl. Data Tables {\bf 59}, 185 (1995).

\bibitem{buchinger} F. Buchinger and J. M. Pearson,  
Phys. Rev. C {\bf 72}, 057305 (2005).

\bibitem{iimura_08} H. Iimura and F. Buchinger, Phys. Rev. C {\bf 78}, 067301 (2008).

\bibitem{iimura_15} Hideki Iimura, Peter M\"oller, Takatoshi Ichikawa, 
Hiroyuki Sagawa, and Akira Iwamoto, 
JPS Conf. Proc. {\bf 6}, 030102 (2015).

\bibitem{lalazissis} G. A. Lalazissis, S. Raman, and P. Ring, 
At. Data Nucl. Data Tables {\bf 71}, 1 (1999).

\bibitem{long} W. Long, J. Meng, N. Van Giai, and S. G. Zhou, Phys. Rev. C
{\bf 69}, 034319 (2004).

\bibitem{geng_05} Lisheng Geng, Hiroshi Toki, and Jie Meng, 
Prog. Theor. Phys. {\bf 113}, 785 (2005).

\bibitem{libert} J. Libert, B. Roussiere, and J. Sauvage, 
Nucl. Phys. A {\bf 786}, 47 (2007).

\bibitem{delaroche} J.-P. Delaroche, M. Girod, J. Libert, H. Goutte, S. Hilaire, 
S. P\'eru, N. Pillet, and G. F. Bertsch, Phys. Rev. C {\bf 81}, 014303 (2010).

\bibitem{goriely} S. Goriely, N. Chamel, and J. M. Pearson, Phys. Rev. 
C {\bf 88}, 024308 (2013);\\
http://www-astro.ulb.ac.be/bruslib/nucdata/hfb24-dat

\bibitem{skin} P. Sarriguren, M. K. Gaidarov, E. Moya de Guerra, and A. N. Antonov,
Phys. Rev. C {\bf 76}, 044322 (2007).

\bibitem{gaidarov1} M. K. Gaidarov, A. N. Antonov, P. Sarriguren, and E. Moya de Guerra,
Phys. Rev. C {\bf 84}, 034316 (2011); {\it ibid.} {\bf 85}, 064319 (2012).

\bibitem{gaidarov2} M. K. Gaidarov, P. Sarriguren, A. N. Antonov, and E. Moya de Guerra,
Phys. Rev. C {\bf 89}, 064301 (2014).


\bibitem{negele} J. W. Negele, Phys. Rev. C {\bf 1}, 1260 (1970).

\bibitem{bertozzi} W. Bertozzi, J. Friar, J. Heisenberg, and J. W. Negele, 
Phys. Lett. B {\bf 41}, 408 (1972).

\bibitem{sick03} I. Sick, Phys. Lett. B{\bf 576}, 62 (2003).

\bibitem{gentile11} T. R. Gentile and C. B. Crawford,
Phys. Rev. C {\bf 83}, 055203 (2011).

\bibitem{goriely_16} S. Goriely, N. Chamel, and J. M. Pearson, 
Phys. Rev. C {\bf 93}, 034337 (2016).

\bibitem{pillet} N. Pillet and S. Hilaire, Eur. Phys. J. A {\bf 53}, 193 (2017).


\end{thebibliography}
\end{document}